# Rashba field in GaN

A. Wolos,[1] Z. Wilamowski,[1,2] C. Skierbiszewski,[3] A. Drabinska,[4]
B. Lucznik,[3] I. Grzegory,[3] and S. Porowski[3]

[1]*Institute of Physics Polish Academy of Sciences,*
*Al. Lotnikow 32/46, 02-668 Warsaw, Poland.*
[2]*Faculty of Mathematics and Computer Sciences, University of Warmia and Mazury,*
*ul. Zolnierska 14, 10-561 Olsztyn, Poland*
[3]*Institute of High Pressure Physics of the Polish Academy of Sciences, Unipress,*
*ul. Sokolowska 29/37, 01-142 Warsaw, Poland*
[4]*Institue of Experimental Physics, Faculty of Physics, University of Warsaw,*
*ul. Hoza 69, 00-681 Warsaw, Poland*

We discuss problem of Rashba field in bulk GaN and in GaN/$Al_xGa_{1-x}N$ two-dimensional electron gas, basing on results of X-band microwave resonance experiments. We point at large difference in spin-orbit coupling between bulk material and heterostructures. We observe coupled plasmon-cyclotron resonance from the two-dimensional electron gas, but no spin resonance, being consistent with large zero-field spin splitting due to the Rashba field reported in literature. In contrast, small anisotropy of g-factor of GaN effective mass donors indicates rather weak Rashba spin-orbit coupling in bulk material, not exceed 400 Gauss, $\alpha_{BIA} < 4\times10^{-13}$ eVcm. Furthermore, we observe new kind of electron spin resonance in GaN, which we attribute to surface electron accumulation layer. We conclude that the sizable Rashba field in GaN/$Al_xGa_{1-x}N$ heterostructures originates from properties of the interface.

## 1. INTRODUCTION

Spin splitting of conduction band in bulk GaN and in GaN-based heterostructures has attracted recently considerable interest due to expected long electron spin coherence times in these materials.[1,2] Long spin relaxation times are highly desired for spintronic applications to maintain information about spin while a spin current travels through a semiconductor. Spin relaxation times depend crucially on the magnitude of spin-orbit interactions, being the longer the weaker is the coupling.[3] Gallium nitride is expected to show weak spin-orbit coupling as both Ga and particularly N are rather light elements. Indeed, valence band spin-orbit splitting in bulk GaN is about 20 times weaker than in GaAs ($\Delta_{GaN} = 0.016$ eV, $\Delta_{GaAs} = 0.35$ eV) and more than two times weaker than in Si ($\Delta_{Si} = 0.044$ eV).[4] Low-temperature spin lifetimes in bulk GaN have been demonstrated to range up to about 20 ns, despite high dislocation density.[1] On the other hand GaN is a polar semiconductor showing effects of spontaneous and piezoelectric polarization resulting in strong electric fields. Any electric field acts on the spin of a moving electron due to spin-orbit coupling mechanism. It cannot be thus excluded that the polarization-induced electric fields can enhance the magnitude of conduction band spin splitting. In GaN/$Al_xGa_{1-x}N$ heterostructures grown along GaN c-axis, with polarization-induced two-dimensional electron gas (2DEG) at the interface, the spin-orbit Rashba effect has been demonstrated to be surprisingly large.[5] The Rashba spin-orbit coupling parameter $\alpha$ has been shown to equal between $5.5 - 10 \times 10^{-11}$ eVcm with the resulting low-temperature spin scattering times being of the order of a ps.[6,7,8,9] For comparison, in modulation doped Si/SiGe quantum wells $\alpha$ is two orders of magnitude weaker, equal to $5\times10^{-13}$ eVcm, with spin scattering times ranging up to a microsecond.[10]

Spontaneous and piezoelectric polarization effects lead to accumulation of conduction electrons or holes at the interface of nitride heterostructures or at the surface of GaN itself. Macroscopic, polarization-induced electric fields can reach up to $7\times10^6$ V/cm in GaN-based heterostructures. Two-dimensional electron gas can be formed spontaneously at the GaN/$Al_xGa_{1-x}N$ interface without additional doping, owing only to a large difference in polarization charge between GaN and $Al_xGa_{1-x}N$.[11] At the nitrogen face of GaN, polarization charge causes band bending allowing formation of electron accumulation layer.[12] In this communication we address the problem of the Rashba-type spin-orbit interaction in both bulk GaN and in two-dimensional electron gas formed either at the GaN/$Al_xGa_{1-x}N$ interface or at the GaN surface.

In bulk semiconductors or semiconductor heterostructures lacking inversion symmetry, splitting of the conduction band occurs even without an external magnetic field. This effect due to bulk inversion asymmetry of a crystal (BIA) has been studied by Dresselhaus in 1955[13] and Rashba in 1960.[14] Dresselhaus have noticed that in crystals with $T_d$ symmetry there is a spin splitting of the conduction band, which is cubic in electron **k**-vector. In uniaxial crystals, the splitting linear in **k**-vector appears.[14] Bychkov and Rashba in 1984 have discussed another type of linear spin-splitting, which is occurring in quantum wells or superlattices due to the structure-induced asymmetry (SIA).[15] The dependence of spin splitting on electron momentum, **k**, can be determined to large extend by symmetry considerations. Neglecting Dresselhaus term proportional to $k^3$, the effective spin Hamiltonian describing splitting of the conduction band in GaN has the form:[16,17,18]

$$\mathscr{H} = \alpha(k_y\sigma_x - k_x\sigma_y) \ . \qquad \text{Eq. 1}$$

Spin splitting is expressed here as a linear function of electron **k**-vector, with $k_i$ being the latter's components. $\sigma_i$ denotes Pauli matrices. $\alpha = \alpha_{BIA} + \alpha_{SIA}$ is the Rashba spin-orbit coupling parameter, being a sum of bulk inversion asymmetry, $\alpha_{BIA}$, and structure inversion asymmetry, $\alpha_{SIA}$, components, respectively. Due to the form of the spin Hamiltonian, it is impossible to distinguish experimentally between BIA and SIA contributions in gallium nitride-based heterostructures.

The spin-orbit interaction expressed by Eq.1 is equivalent to appearing of the effective magnetic field, the Rashba field, which is oriented in-plane of the 2DEG and perpendicular to the electron **k**-vector. The magnitude of the Rashba field is proportional to the magnitude of the **k**-vector, k:

$$B_R = \frac{2\alpha k}{g\mu_B} \ . \qquad \text{Eq. 2}$$

Here, g is electron g-factor and $\mu_B$ is Bohr magneton. In a case of bulk GaN there is, of course, no $\alpha_{SIA}$ contribution. The whole Rashba field originates then from the symmetry of wurtzite structure.

The relation of Rashba parameter $\alpha$ to electric field in a heterostructure has remained controversial for many years. De Andrada e Silva *et al.* have shown that Rashba parameter can be derived from $\mathbf{k}\cdot\mathbf{p}$ model. In a uniform electric field, E, and neglecting interface effects, the structure-related $\alpha_{SIA}$ parameter equals to $\alpha_{SIA} = \alpha_0 E$.[19] The proportionality parameter, $\alpha_0$, is a constant characteristic for the particular semiconductor, dependent on its energy gap and spin-orbit splitting of the valence band. Pikus and Pikus have noticed that the electric field, E, should be replaced by its average value for more general cases. This, however, gives $\alpha_{SIA} = 0$ if the effective mass approximation would have been valid throughout the entire well, including the barriers.[20] Following the problem, Pfeffer and Zawadzki have calculated spin splitting for $In_{0.53}Ga_{0.47}As/In_{0.52}Al_{0.48}As$ heterostructure, including in their $\mathbf{k}\cdot\mathbf{p}$ Hamiltonian effects of potential discontinuity at the interface.[21] They have concluded that the average electric field in the heterostructure contributes only to 3% of the total SIA spin splitting, while the dominant contribution comes from the interface. The latter depends only on valence bands offsets and on the envelope function at the interface. The authors have included both effects of BIA and SIA to their calculation. They have shown that major contribution to total spin splitting comes from the SIA. It is now agreed that effects of interface play important role in spin-orbit splitting of the Rashba type in asymmetric heterostructures.

Another approach to Rashba splitting, based on first-principles, relativistic local density calculations, has been proposed by Majewski.[17] The author has investigated influence of spontaneous and piezoelectric polarization in AlN/GaN superlattices on the magnitude of Rashba parameter. It has been calculated that GaN is characterized by rather high $\alpha_{BIA}$ parameter equal to $9\times10^{-11}$ eVcm, owing to spontaneous polarization of the bulk material (This value is already as high as Rashba parameter determined for GaN/$Al_xGa_{1-x}N$ heterostructures in weak antilocalization experiments). In unstrained wells grown on GaN substrate, the calculations show that $\alpha$ is slightly dependent on the width of the quantum well, equal to between $0.4\times10^{-11}$ and $0.8\times10^{-11}$ eVcm. It has been concluded that the electric field at the interface originating from difference in electric polarization between GaN and AlN counteracts effects of spontaneous polarization of the bulk GaN, leading to reduction of the Rashba parameter. Moreover, the parameter $\alpha$ have appeared very sensitive to strain due to the lattice mismatch to the substrate. In particular, compressive biaxial strain causes considerable increase of the conduction band spin splitting.

In this communication we address the issue of the Rashba field in bulk GaN and in GaN/$Al_xGa_{1-x}N$ heterostructures. Microwave resonance measurements in X-band (f=9.5 GHz) were performed both on bulk GaN:Si crystals grown by Hydride Vapor-Phase Epitaxy (HVPE) and on the GaN/$Al_xGa_{1-x}N$ two-dimensional electron gas. We investigate Rashba field acting on electrons in bulk GaN and evaluate value of the $\alpha_{BIA}$ parameter, basing on the analysis of g-factor anisotropy of GaN effective mass donors. We use formalism of the Rashba field successfully applied earlier to electron spin resonance in asymmetric Si/SiGe quantum wells.[10,22,23] In a case of GaN/$Al_xGa_{1-x}N$, we conclude that lack of spin resonance of the 2DEG in X-band is consistent with large zero-field spin splitting due to the Rashba field reported in literature. Besides well-known donor signal, we present a new kind of spin resonance in bulk GaN, which we attribute to electrons accumulated at the GaN surface. Both electrons localized on donors and these at the GaN surface undergo action of rather weak Rashba field, not exceeding a few hundred Gauss in both cases This is in contrast to $GaN/Al_xGa_{1-x}N$ heterostructures where the Rashba field is strong. We discuss possible explanations of the observed effects.

## 2. SAMPLES AND EXPERIMENTAL DETAILS

$GaN/Al_xGa_{1-x}N$ heterostructures were grown by plasma-assisted molecular beam epitaxy on GaN single crystals. The substrates were obtained by high-pressure technique. The wurtzite c-axis was perpendicular to the heterostructure interface. The 2DEG concentration was typically of the order of $2 \times 10^{12}$ $cm^2$, while the Hall mobility ranged between 80 000 and 5 000 $cm^2/(Vs)$. The aluminum content x was set to 0.09.[24]

GaN bulk crystals were deposited by Hydride Vapor-Phase Epitaxy on GaN high pressure-grown seeds.[25] The crystals were n-type, doped with Si. Donor concentration was equal to $2 \times 10^{18}$ $cm^{-3}$. Samples were cut to obtain lateral dimension equal to about $4 \times 4$ $mm^2$ to fit to the sample holder. The thickness was equal to 0.2 mm. GaN c-axis was perpendicular to the sample plane.

Electron spin resonance (ESR) was measured using Bruker ESP 300 spectrometer operating in X-band (9.5 GHz). The $TE_{102}$ mode was used, for which in unperturbed microwave cavity the external magnetic field is parallel to the microwave electric field. At the center of the cavity, where a sample should be placed, microwave magnetic field has its maximum while the electric field remains residual. The temperature was varied down to 2.5 K using Oxford continuous-flow cryostat. Magnetic field was calibrated with a standard marker.

## 3. MAGNETOPLASMA RESONANCE IN $GaN/Al_xGa_{1-x}N$ AND LACK OF SPIN RESONANCE

Microwave resonance technique has been usually associated with measurements of spin resonance spectra, whereas in addition to these, semiconducting samples placed in a microwave cavity of a standard ESR spectrometer show also spectral features related to the electric conductivity. These can be either Shubnikov-de Haas oscillations, pure cyclotron resonance or a magnetoplasma resonance, which is a coupled plasmon-cyclotron excitation.[26] Figure 1 shows edge magnetoplasma resonance recorded for $GaN/Al_xGa_{1-x}N$ samples. The signal is sensitive only to perpendicular to the sample plane component of the applied magnetic field, which is characteristic for 2DEGs. The resonance position of two lowest magnetoplasma branches for the disc-shaped sample has been calculated by Allen *et al*:[27]

$$\omega_{res}^{\pm} = \pm\frac{\omega_c}{2} + \sqrt{\omega_p^2 + \left(\frac{\omega_c}{2}\right)^2}\ , \qquad \text{Eq. 3}$$

where $\omega_{res}^{+}$ denotes the upper cyclotron-like branch and $\omega_{res}^{-}$ is the lower edge mode. $\omega_c$ and $\omega_p$ are cyclotron and plasma frequencies, respectively. In a case of a 2DEG, plasma frequency is dependant on sample lateral dimension. Sheet electron concentration ~2 $\times 10^{12}$ $cm^2$, typical for $GaN/Al_xGa_{1-x}N$ heterostructures, together with sample lateral diameter of a few mm provide magnetoplasmaplasma frequency suitable to meet X-band resonance condition. Higher electron concentration, exceeding $10^{13}$ $cm^2$, makes the edge resonance to appear at higher magnetic field. This is (i) beyond a range of typical X-band ESR spectrometers and (ii) the amplitude of the resonance weakens strongly with increasing magnetic field.

Combining the theory of magnetoplasma excitations with the Drude model of momentum relaxation one can evaluate both plasma frequency and mobility of the 2DEG from the magnetoplasma spectrum. The parameters obtained for $GaN/Al_xGa_{1-x}N$ samples by following the fitting procedure described in Ref. 26 are compared to Hall data in Fig. 1. The mobility determined from the magnetoplasma resonance is for high-mobility samples about two times higher than obtained by Hall experiment. This is most probably due to high-frequency character of the experiment, which measures local mobility but not macroscopic mobility through a whole sample. The resulting "magnetoplasma mobility" is higher than static Hall mobility. The magnetoplasma resonance confirms presence of high-quality two-dimensional electron gas in our $GaN/Al_xGa_{1-x}N$ samples.

Despite that we measured many different samples with various mobility and sheet electron concentration we did not record any signal which could be attributed to spin resonance of 2D electrons. There is also lack of this kind of spectra in literature reports. Spin resonances due to 2DEG has been observed in X-band in Si/SiGe quantum wells,[10,22,23] AlAs quantum well,[28] but never in $GaN/Al_xGa_{1-x}N$ heterostructures.[29]

Lack of the spin resonance in $GaN/Al_xGa_{1-x}N$ appears less surprising when taking into account the Rashba zero-field splitting of the conduction band. Besides a series of weak antilocalization experiments, the zero-field spin splitting has been studied by investigations of beatings of Shubnikov-de Haas oscillations,[30] or conductance features of quantum point contact.[31] It has been reported to reach between 9 meV (Ref. 30) and 0.4 meV (Ref. 31). From weak antilocalization experiments the Rashba spin-orbit parameter has been evaluated to be equal to about $\alpha = 6 \times 10^{-13}$ eVcm, which together with a typical sheet electron concentration ($n_{2D}$ = 2 ×$10^{12}$ $cm^{-2}$ is assumed) corresponds to the Rashba field of about $B_R$ = 40 000 Gauss, Eq. 2. Such a large spin-orbit field creates zero-field splitting equal to $\Delta E_{S\text{-}O} = 2\alpha k =$ 0.4 meV. This energy falls far beyond the resonance condition of the X-band (9.5 GHz = 40 $\mu$eV ). It remains established that the zero-field splitting in GaN/$Al_xGa_{1-x}N$ grown along GaN c-axis is sizable,[5] which explains lack of the spin resonance in X-band.

## 4. SPIN RESONANCE IN BULK GaN:Si AND LACK OF MAGNETOPLASMA RESONANCE

Figure 2 shows electron spin resonance spectra of bulk GaN:Si samples. Two lines having similar g-factor and its anisotropy are visible, a broad slightly Dysonian line, and a narrow line having pure Lorentzian shape. Figure 2 shows decomposition of the measured signal into the two components. The broad line is a well-known resonance due to effective mass donors, in our case Si, originally described by Carlos and coworkers.[32] Properties of this resonance in HVPE-grown samples have been discussed in details in our previous communication.[33] The Dysonian lineshape of the broad resonance appears due to conducting character of samples doped with 2 ×$10^{18}$ of Si donors per $cm^3$. From the lineshape one can evaluate skin depth for microwave penetration, which equals to 0.35 mm at the temperature T = 2.5 K and 0.17 mm at T = 16 K. The decrease of the skin depth with increasing temperature is related to thermal activation of the conductivity, with the activation energy equal to about 0.6 meV. The g-factor of the broad line is anisotropic with $g_{\parallel} = 1.9509$ and $g_{\perp} =$ 1.9480, similar to values reported in Ref. 32, Fig. 3(a). The linewidth is also slightly anisotropic, equal to 15 and 22 Gauss at T = 2.5 K, for the external magnetic field oriented parallel and perpendicular to the GaN c-axis, respectively, Fig. 3(b). The linewidth increases when increasing the temperature, Fig. 4(a). Main effects leading to line broadening are interactions with GaN lattice phonons at high temperatures and excitations to double-occupied Si donor states. The latter mechanism shortens electron lifetime in the single-occupied Si donor band which imposes a limit to the spin relaxation time.[33]

The narrow line is well visible on the background of Si donor resonance only in a narrow temperature range, between 10 K and 30 K. Its g-factor is very close to that of Si donors. It equals to $g_{\parallel} = 1.9512$ and $g_{\perp} =$ 1.9481. Other properties are, however, very different from the donor signal. The line has width equal to about 3 Gauss independent on temperature, Fig. 4(a). Temperature dependence of the signal amplitude is also unusual.

While the donor signal shows 1/T dependence characteristic for Curie paramagnetism, the amplitude of the narrow line can be rather described by a $1/T^{3.3}$ relation, Fig. 4(b). In addition, the signal amplitude shows angular anisotropy, differing by a factor of two for two perpendicular orientations of the sample inside resonance cavity. Pure magnetic dipole transitions characteristic for donor centers should not depend on the orientation. In contrast, electric-dipole resonance or a so-called current-induced spin resonance, where electric component of the microwave field induces spin resonance through respective mechanisms, is known to be crucially depend on the geometry.[34] These kinds of excitations are characteristic for conducting electrons, in particular these of the two-dimensional electron gas.

To summarize, a following properties can be concluded from the narrow resonance line: (i) the resonance originates from GaN effective mass electrons, as indicated by the anisotropy of the g-factor; (ii) pure Lorentzian lineshape suggests, that the resonance results from centers distributed close to the sample surface, which are not sensitive to skin effect; (iii) small linewidth is characteristic for delocalized electrons, for which spin relaxation time is modulated by a characteristic correlation time, related, *e.g.* to momentum scattering time (Dyakonov-Perel mechanism); (iv) the dependence of the signal amplitude on the geometry in the resonance cavity points at electric field-driven spin resonance, characteristic for conducting electrons. The electron system which possesses such properties can be attributed to the *surface accumulation layer*.

It has been shown that spontaneous and piezoelectric polarization leads to appearing of positive bound charge at the nitrogen-face of GaN. The surface charge induces downward bending of the GaN conduction band, which can lead to accumulation of conduction electrons at the surface. The sheet concentration of the bound polarization charge is of the order of $2\times10^{13}$ $cm^{-2}$.[12]

We did not observe magnetoplasma resonance due to electron surface accumulation layer in X-band up to the magnetic field of 10 000 Gauss. The resonance can be easily observed only when microwave frequency is of the same order as plasma frequency. For the plasma frequency of 2D electrons having sheet concentration $2 \times 10^{13}$ $cm^{-2}$, which corresponds to the concentration of the surface polarization charge in GaN, the resonance falls beyond range of our spectrometer, Eq. 3.

## 5. g-FACTOR ANISOTROPY IN GaN

Electron spin resonance is a natural way to study spin-orbit interactions. Rashba type of spin splitting has been systematically studied by ESR in Si/SiGe quantum wells. It turned out that Rashba field, which is of the order of a few hundred Gauss in Si/SiGe, influences both the g-factor and the linewidth of the resonance signal. The g-factor of conduction electrons, which is isotropic in bulk Si, gains its anisotropy due to in-plane oriented Rashba field appearing thanks to asymmetric structure of the quantum well. The anisotropy is the more pronounced the higher is electron concentration, so the larger is electron **k**-vector at the Fermi level. The narrow width of spin resonance can be described by Dyakonov-Perel relaxation. Here the dependence of the Rashba field on electron **k**-vector causes spread in resonance frequency, which additionally is averaged by momentum scattering events. For a review of these effects, see *e.g.* Ref. 22.

In a case of wurtzite GaN, the g-factor of shallow donors has been analyzed so far within a formalism of 7-band $\mathbf{k}\cdot\mathbf{p}$ model.[35] The authors have concluded that the anisotropy of the g-factor is caused by spin-orbit and crystal-field splittings of higher conduction bands. Unfortunately, no detailed information on the band structure of higher conduction bands in GaN is known, so the model parameters have been only estimated. It is also lacking of a calculation accounting for wurtzite symmetry and remote band effects, which simultaneously evaluates the magnitude of the linear in **k**-vector spin splitting of the Rashba-type.

In this chapter we will estimate experimentally the upper limit for the Rashba filed in GaN basing on the analysis of the g-factor anisotropy of effective mass electrons, either localized on Si donors or these accumulated at the GaN surface. We will use formalism applied earlier to Si/SiGe. Formulas presented below originate from symmetry considerations taking into account the particular form of the effective spin Hamiltonian, Eq. 1.[10,22,22]

The Rashba field adds up to the external applied magnetic field shifting the spin resonance. At sufficiently low temperatures only electrons in the vicinity of the Fermi vector participate in the resonance. In 2D structures with dominating Rashba spin-orbit interaction, the resonance field is expressed in the form:[10,22,23]

$$B_{res} = B_0 - \frac{B_R^2}{4B_0}(1+\cos^2\theta)\,. \qquad \text{Eq.4}$$

Equation 4 has been obtained by averaging the contribution from all electrons at the Fermi circle. Here, $B_{res}$ is the observed resonance field and $B_0$ is the resonance magnetic field without the Rashba field. The anisotropy of the resonance field is expressed here as a function of a square of the Rashba field for the Fermi vector:

$$B_R^2 = \left( \frac{2\alpha\ k_F}{g\mu_B} \right)^2 . \qquad \text{Eq.5}$$

Applying Eq. 4 to the narrow signal accounted in Chapter 4 for electrons accumulated at the GaN surface, Fig. 3(a), one gets a value of the Rashba field $\sqrt{B_R^2}$ = 277 Gauss. This is an upper limit for the Rashba field, when all other contributions to the g-factor anisotropy are neglected. It can be already seen that this value is drastically smaller than $B_R$ = 40 000 Gauss evaluated for $GaN/Al_xGa_{1-x}N$ heterostructures.

Similar analysis of the g-factor anisotropy can be performed for Si donors. The g-factors of shallow donors should be close to that of the conduction band. With this assumption, $\mathbf{k}\cdot\mathbf{p}$ models are usually applied to shallow donor resonances. The evaluated Rashba field will give us information about the magnitude of bulk inversion asymmetry contribution as there is no structure-induced inversion asymmetry in this case. We will perform a following procedure. Eq.4 should be modified to the form:

$$B_{res} = B_0 - \frac{B_R^2}{8B_0}(1+\cos^2\theta), \qquad \text{Eq.6}$$

to account for 3D space. $B_R^2$ is given by Eq.5 like in a 2D case, with $k_F$ being a value of the **k**-vector at the Fermi sphere. The Rashba field determined in this way equals to $B_R$ = 380 Gauss. To evaluate the Rashba coefficient it is necessary to know the value of the Fermi vector. In the discussed case we analyze electrons localized on shallow donors. Their wavefunction is constructed from wavefunctions of conduction band states. Here, we will relate the magnitude of their **k**-vector to the localization radius on the donor. We assume $k_F$, necessary to determine parameter $\alpha$ in Eq.5, to be equal to $k_F = 2\pi/\lambda$ and $\lambda = 4R$. Where R is electron localization radius on the Si donor, equal to about 2.8 nm ($R = \sqrt{\hbar^2/(2m^*E_I)}$, $E_I$ is ionization energy of Si donor), and $\lambda$ denotes a wavelength. This assumption leads to $k = 6 \times 10^6$ $cm^{-1}$, which gives $\alpha_{BIA} = 4\times10^{-13}$ eVcm. Again, like in a case of surface accumulation layer, the determined value of Rashba spin-orbit parameter is only an upper limit. Accounting for crystal-field splitting due to wurtzite structure can lead to diminishing of the value of $\alpha$.

## 6. DISCUSSION AND CONCLUSIONS

Even when a sizable Rashba field is observed in $GaN/Al_xGa_{1-x}N$ heterostructures grown along GaN c-axis ($B_R$ = 40 000 Gauss), neither electrons localized on effective mass donors nor these accumulated at the GaN surface do not feel the action of such a large spin-orbit field. We have argued in previous chapters that the Rashba field cannot exceed about 400 Gauss in bulk GaN, with the upper limit for the bulk inversion asymmetry parameter $\alpha_{BIA}$ equal to $4\times10^{-13}$ eVcm. In bulk GaN filled with shallow donors or with the 2DEG accumulated at the surface the Rashba field is weak, which is in large contrast to $GaN/Al_xGa_{1-x}N$ heterostructures. Similar situation has been observed for $In_{0.53}Ga_{0.47}As/In_{0.52}Al_{0.48}As$ quantum well discussed in the introduction, for which case the $\alpha_{SIA}$ has been calculated to dominate strongly over $\alpha_{BIA}$.[21] The contribution of the structure-induced asymmetry to the spin splitting have originated in that case from the properties of the interface, in particular from valence band parameters on both sides of the interface and on the envelope function at the interface. In a case of $GaN/Al_xGa_{1-x}N$ heterostructures, one cannot totally omit the role of electric fields in the well, especially that spontaneous and piezoelectric polarization effects are particularly strong in nitrides.[17] Electric field can act on the envelope wave function at the interface, influencing the magnitude of spin splitting by changing electron penetration depth into the barrier. This mechanism has been proposed to explain effects of gate voltage on the Rashba splitting in InAs quantum well.[36]

We would like to stress, that in a case of nitride heterostructures it is not a macroscopic electric field which is responsible for large Rashba field but rather a field originating from microscopic properties of the interface alone. To support this statement it is enough to recall results of weak antilocalization experiments.[6,7,8,9] Some of them have been performed on gated samples allowing modulation of sheet electron concentration by applying electric field to the heterostructure. The electric field modified the Rashba spin splitting, but no effects of the applied voltage on the Rashba parameter $\alpha$ have been demonstrated. Theoretical calculations presented by Majewski, Ref.17 show also that the Rashba splitting remains rather robust to the external bias. We performed also some experiments on gated, modulation doped $GaN/Al_xGa_{1-x}N$ quantum wells applying voltage which allowed either compensation or enhancement of macroscopic electric field in the quantum well. The effect of the applied voltage was monitored by electroreflectance measurements. We could obtain up to the full compensation of built-in electric field in the quantum well. These experiments, however did not lead us to suppression of the Rashba field to a value allowing observation of spin resonance in conditions offered by X-band ESR spectrometer. This result indicates again that the Rashba field cannot be efficiently tuned by modulation of electric field in GaN quantum wells. To summarize, both theoretical and experimental data point out at the importance of the interface in a problem of Rashba field in $GaN/Al_xGa_{1-x}N$ heterostructures.

At this point, we should comment on the value of parameter $\alpha$ calculated by Majewski in Ref. 17 within first-principles, relativistic local density approach. The Rashba parameter given in Ref. 17 equals to $\alpha_{BIA} = 9\times10^{-11}$ eVcm, with its large value originating from spontaneous polarization of bulk material. In this paper we have shown that electrons localized on donors in n-type GaN do not feel the Rashba field higher than 400 Gauss, which corresponds to $\alpha_{BIA} < 4\times10^{-13}$ eVcm. The discrepancy between calculated and measured values can be accounted for the fact, that in a real crystal polarization effects are screened by presence of conducting electrons. This leads apparently to reduction of the value of the Rashba parameter. $\alpha_{BIA}$ given in Ref. 17 is of the same order as the value determined from weak antilocalization experiments for 2DEG in $GaN/Al_xGa_{1-x}N$. We have argued that such a sizable Rashba field as it is observed in the heterostructures results in large zero-field spin splitting, disabling observation of spin resonance in X-band. The fact that we can record spin resonance due to effective mass donors in GaN with its rather inconsiderable anisotropy, testifies for weak Rashba field in this system.

In this communication, we have reported for the first time spin resonance originating from the electron surface accumulation layer in GaN. Further studies on that signal are required to determine basic electric parameters like sheet electron concentration and mobility of the 2DEG. The mechanisms responsible for spin excitation are to be clarified as well. Angular anisotropy of the signal amplitude mentioned in Chapter 4 indicates that the resonance transition is driven by microwave electric field. However, to investigate specific selection rules for spin excitation, in particular the dependence of the signal amplitude on the geometry of electric field, it is necessary to perform experiments in a resonance cavity allowing placing a sample in the maximum of the electric field. In a case of unperturbed $TE_{102}$ cavity used in this experiment, the electric field equals to zero in the center where the sample should be placed. The fact that we could still observe resonances induced by the electric field, the magnetoplasma resonance described in Chapter 3 included here as well, means that either the cavity was significantly perturbed by introducing a sample, or due to finite dimension the sample went out of the very center of the cavity to the non-vanishing electric field. In any case, it is not possible to control these effects in a standard $TE_{102}$ setup, thus different geometry of the resonance cavity is here necessary to clarify these effects.

The mobility of the surface 2DEG we expect rather sizable. In Dyakonov-Perel model of spin relaxation, which is the most common spin relaxation mechanism, the dependence of the resonance frequency on electron **k**-vector leads to broadening of the resonance line. The original spread in resonance frequency is motionally averaged, owing to the fact that momentum relaxation is usually much faster than spin relaxation. In a result the resonance line is of the Lorentzian shape with a width of:

$$\Delta\omega = \Omega^2\tau . \qquad \text{Eq. 7}$$

Where, $\Delta\omega$ is a linewidth expressed in frequency units, $\Omega^2$ is a variance of the distribution of Rashba frequency, $\hbar\Omega = g\mu_B B_R$, $\tau$ is momentum scattering time. Applying Eq. 7 to the ESR signal from electrons accumulated at the GaN surface, having the width equal to $\Delta\omega = \gamma\, 3$ Gauss ($\gamma$ is electron gyromagnetic ratio) and Rashba field equal to $B_R$ = 277 Gauss, one obtains $\tau = 2\times10^{-12}$ s corresponding to the mobility equal to about 20 000 $cm^2/(Vs)$. Keeping in mind that 277 Gauss is only the upper limit for the Rashba field one can expect that the mobility in the surface accumulation layer can reach even higher values.

Magnetoplasma resonance could help in determination of the sheet electron concentration of the surface 2DEG. To meet the resonance condition for high-concentration sample one needs, however, a frequency higher than 9.5 GHz offered by the X-band spectrometer. 40 GHz of the Q-band would be suitable to observe the resonance at low magnetic field with a reasonable amplitude.

Summarizing, we have investigated problem of Rashba field in bulk GaN:Si and in $GaN/Al_xGa_{1-x}N$ 2DEG. Lack of the X-band ESR in $GaN/Al_xGa_{1-x}N$ is consistent with large Rashba splitting reported in literature. In contrast, anisotropy of g-factor of GaN effective mass donors indicates that Rashba field does not exceed 400 Gauss in bulk material, with Rashba parameter $\alpha_{BIA}$ being not higher than $4\times10^{-13}$ eVcm. Electrons accumulated at the GaN surface do not feel Rashba field larger than 300 Gauss, neither. We conclude that spin-orbit interactions of Rashba type are weak in bulk GaN, whereas the sizable zero-field splitting observed in $GaN/Al_xGa_{1-x}N$ 2DEG results from the interface.

## 7. ACKNOWLEDGEMENTS

This work has been supported by funds for science, grant numbers: PBZ/MNiSW/07/2006/39 and N N202 1058 33, Poland. We would like to thank K. P. for GaN/$Al_xGa_{1-x}$N quantum wells.

FIGURES:

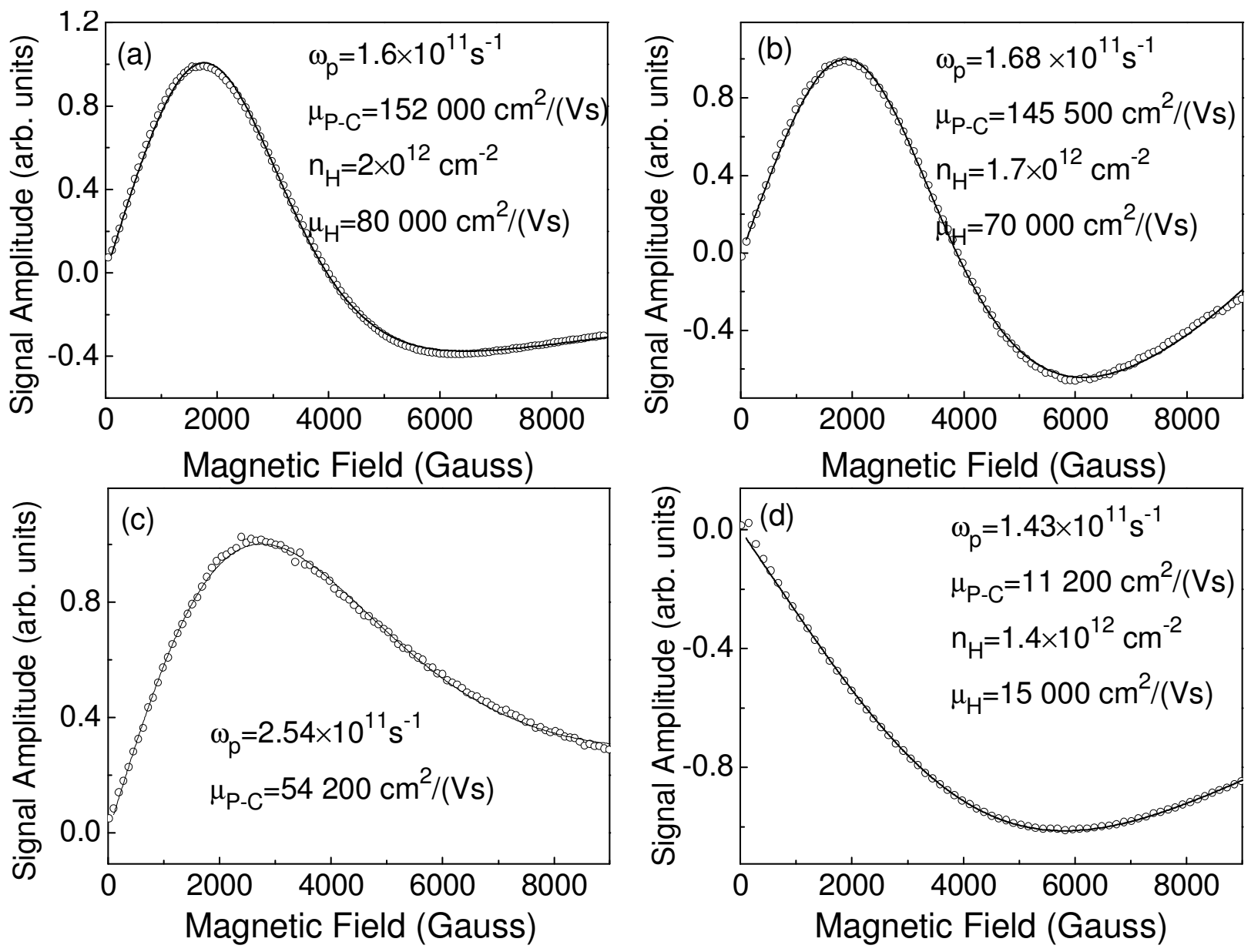

Figure 1(a)-(d). Edge mode of plasmon-cyclotron resonance recorded in X-band for GaN/$Al_{1-x}Ga_x$N samples with various mobility. T = 2.5 K, B || c. Dots are experimental data, solid lines are fitted according to the model of dimension-dependent plasmon-cyclotron coupling and the Drude model of momentum relaxation (Ref. 26). Hall concentration ($n_H$) and Hall mobility ($\mu_H$) of the 2DEGs are indicated in respective figures together with plasma frequency ($\omega_p$) and mobility ($\mu_{P\text{-}C}$) determined from the fit.

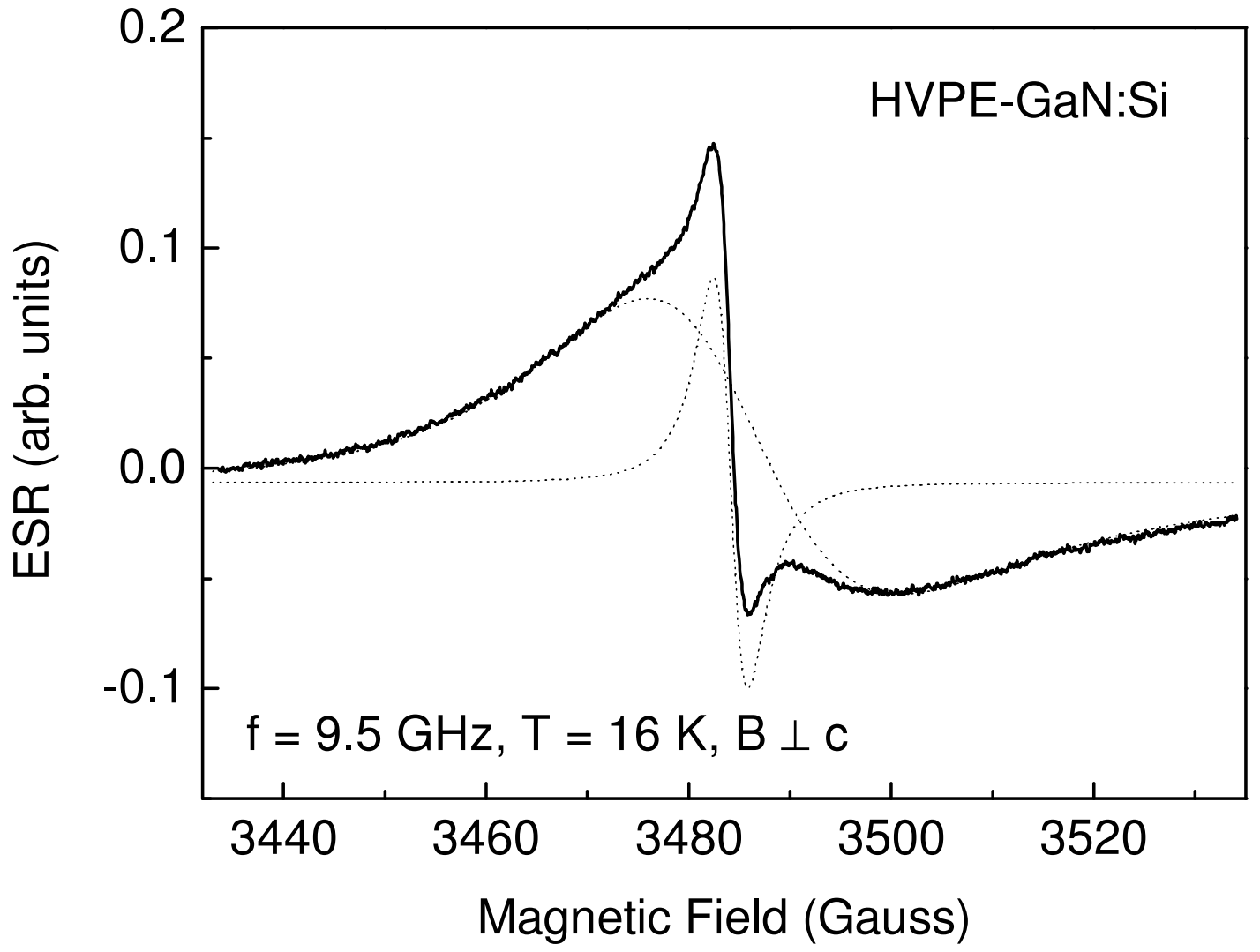

Figure 2. Electron spin resonance in HVPE-grown GaN:Si. Dotted lines show deconvolution into two component lines, slightly Dysonian broad line (HWHM=21 Gauss) and pure Lorenzian narrow line (HWHM=3 Gauss). The broad line is due to Si donors, the narrow one we attribute to effective mass electrons accumulated at the GaN surface.

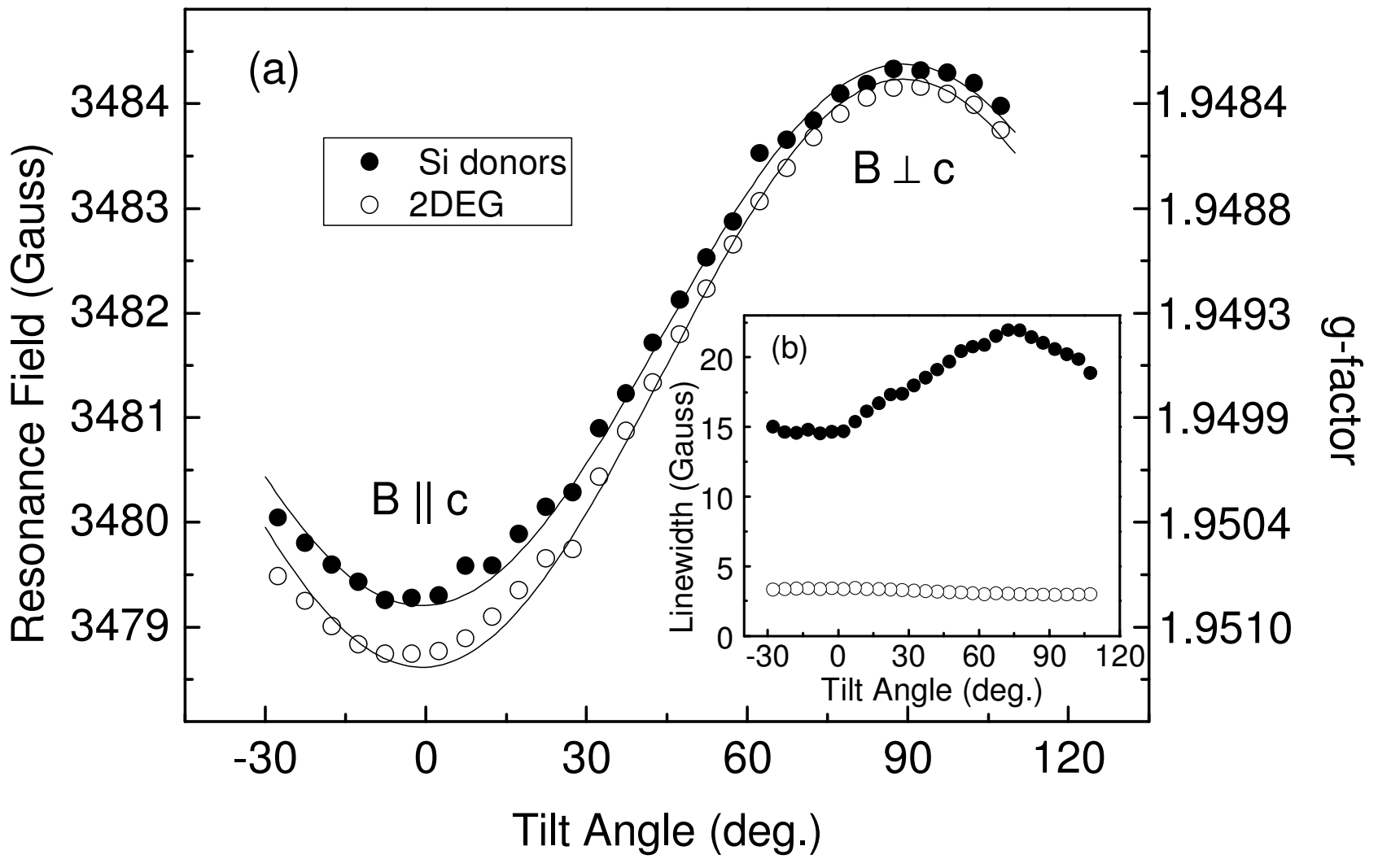

Figure 3. (a) Angular anisotropy of the resonance field for the two lines observed in HVPE-grown GaN:Si. Solid lines are fitted according to the Rashba model. T = 2.5 K. (b) Angular anisotropy of the resonance linewidth. Broad linewidth is characteristic for Si donors, narrow linewidth indicates delocalized states.

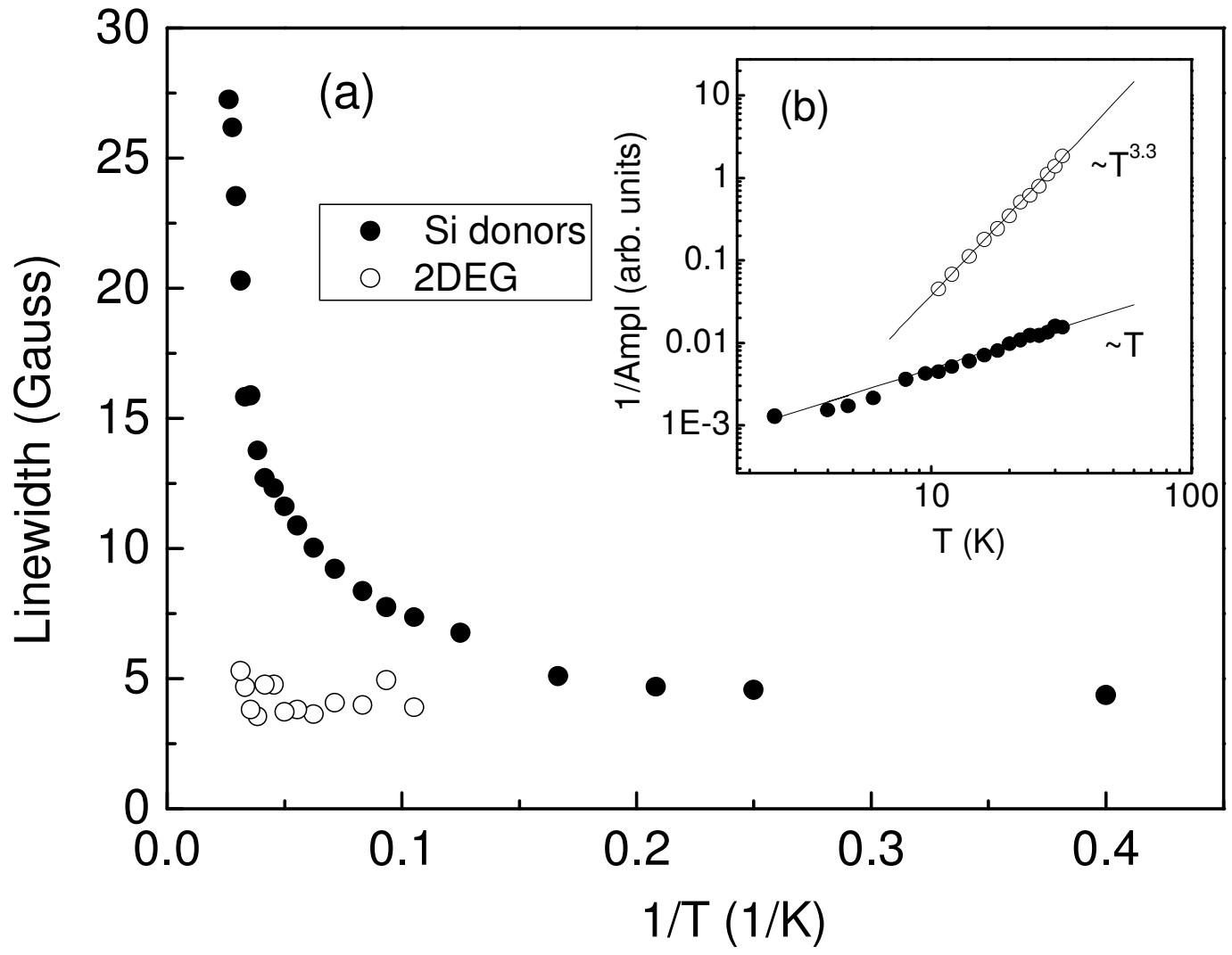

Figure 4. (a) Temperature dependence of the resonance linewidth for Si donors (broad resonance) and electrons at the surface accumulation layer (narrow resonance). Si donor resonance broadens while elevating temperature due to excitations to double-occupied donor band and interactions with lattice phonons. The linewidth of the narrow resonance is not sensitive to temperature in the measured temperature range. (b) Temperature dependence of the amplitude of both signals. Donor signal shows Curie paramagnetism. Narrow signal shows $T^{-3.3}$ dependence of the signal amplitude on temperature.